\documentclass{PoS}

\usepackage{bm}
\usepackage{amsmath}
\usepackage[square,comma,numbers,sort&compress]{natbib}
\usepackage{color}
\usepackage{graphicx}
\usepackage{bbm}

\newcommand*{\La}{\cal{L}}
\newcommand*{\no}{\noindent}
\newcommand*{\bea}{\begin{eqnarray}}
\newcommand*{\eea}{\end{eqnarray}}
\newcommand*{\be}{\begin{equation}}
\newcommand*{\ee}{\end{equation}}

\newcommand*{\nn}{\nonumber}

\title{Correlation Functions and Confinement in Scalar QCD}
\ShortTitle{Scalar QCD}

\author{\speaker{Tajdar Mufti}\thanks{Supported by the DFG graduate school 1523-1 and under grant number MA 3935/5-1}\\
        E-mail: \email{tajdar.mufti@uni-jena.de}}
 \author{{Axel Maas}\thanks{Supported by the DFG under grant number MA 3935/5-1}\\
        E-mail: \email{axelmaas@web.de}\\
\\
        Institute of Theoretical Physics, Friedrich-Schiller-University Jena, Max-Wien-Platz 1, D-07743 Jena, Germany} 

\abstract{The complete knowledge of a theory is encoded in its correlation functions. Thus non-perturbative effects, like confinement in QCD, is necessarily contained in these correlation functions. As a consequence, a number of confinement scenarios make predictions for the behavior of these correlation functions. Non-perturbative methods, like lattice calculations, permit to calculate the correlation functions and test these predictions. To avoid the entanglement with chiral symmetry breaking and the costs of full fermion simulations, here scalar QCD will be used as a role model, as it is expected that confinement operates in the same way. We present results on both two-point functions and three-point functions for the case of two colors and two flavors of scalars, and compare them to the predictions.}

\FullConference{31st International Symposium on Lattice Field Theory - LATTICE 2013\\
		July 29 - August 3, 2013\\
		Mainz, Germany}

\begin{document}

\section{Introduction}

Confinement in the loose sense of the absence of quarks and gluons as free particles \cite{pdg} is a long-standing challenge. In principle, this is resolved by axiomatic field theory \cite{Haag:1992hx}, which shows that no gauge-invariant state with the quantum numbers of quarks and gluons can be defined. However, how this manifests itself in the correlation functions encoding the contents of the theory is far less clear. But this is of central importance for the description of observables using functional methods, where the gauge-dependent correlation functions serve as intermediate steps \cite{Fischer:2006ub,Maas:2011se}. Furthermore, though confinement in the strict sense of a finite string tension \cite{Greensite:2003bk} is absent in QCD, it manifests itself as a residual effect, e.\ g.\ in form of Regge trajectories. How this is arising from the correlation functions in functional methods is also still unclear.

At least in principle, these questions can be answered using lattice calculations, avoiding gauge-dependent intermediate steps. However, continuum methods offer the potential to also describe situations where lattice calculations become impractical, e.\ g.\ when it comes to very different scales. Thus it appears reasonable to pursue both approaches simultaneously, to maximize the methodological reach \cite{Maas:2011se}. This requires to understand how the confining properties are encoded in correlation functions. Lattice calculations can serve as a useful benchmark in special settings for functional methods when it comes to calculating correlation functions at intermediate energies \cite{Maas:2011se}.

The confining properties of QCD are quite involved, and fermion simulations are furthermore very costly. A much simpler and cheaper role model is scalar QCD. It is furthermore free of the complexities induced by chiral symmetry, and the tensorial structures are much simpler.

These advantages come at a cost. It is still unclear whether scalar QCD is a trivial theory. For the following it will just be assumed that if it is, the lattice cutoff removing the triviality has no significant effect on the low-energy properties. This assumption is supported by checking the dependency of the results on the lattice cutoff \cite{Maas:unpublished3}. A second problem is that there is no gauge-invariant distinction between a would-be Higgs phase and a would-be confinement phase of the theory \cite{Fradkin:1978dv}, at least on a finite lattice. However, it turns out that the gauge-invariant spectrum of the theory has quite distinct properties \cite{Maas:2013eh,Jersak:1985nf}. In the would-be QCD phase the lightest excitation is a flavor-singlet $0^{+}$ state, while in the would-be Higgs phase it is a $1^{-}$ flavor triplet, at least for two flavors of scalar quarks. In the cross-over region \cite{Caudy:2007sf}, these states become essentially degenerate. This observation will serve here as an operational definition of what is meant by scalar QCD. Below, only results will be shown where the scalar is a much lighter state than the vector state.

This provides a setup where the two-point and three-point functions of the theory can be calculated on the lattice, and the results will be shown below. This permits particularly to test the predictions for these functions \cite{Fister:2010yw,Fischer:2009tn} and to perform a comparison to the quenched case \cite{Maas:2011yx}. Especially, it permits to check the long-standing idea \cite{Alkofer:2000wg} whether the matter-gluon vertex already yields a (pre-)manifestation of confinement.

\section{Setup}

We consider scalar-Yang-Mils theory with two colors and two flavors, i.\ e.\ the same content as in the Higgs sector of the standard model. This is described by the Lagrangian
\bea
\La&=& -\frac{1}{4}F_{\mu \nu}^{a} F_{a}^{\mu \nu} + (D_{\mu}^{ij}\phi_{j})^{\dagger} (D_{ik}^{\mu}\phi^{k})- \frac{1}{2} m^{2} \phi ^{i\dagger}\phi^{i} - \lambda (\phi ^{i\dagger}\phi^{i})(\phi ^{j\dagger}\phi^{j})\nn\\
D^{ik}_{\mu} &=&\delta^{ik} \partial_{\mu}-igW^{a}_{\mu} t^{ik}_{a}\nn\\
F^{a}_{\mu \nu} &=& \partial_{\mu}W^{a}_{\nu}-\partial_{\nu} W^{a}_{\mu} -g f^{abc} W^{b}_{\mu} W^{c}_{\nu}\nn,
\eea
\no where $\phi$ are the scalar quarks, $W_{\mu}^a $ are the gluons, and $f^{abc}$ and $\tau^a$ are the structure constants and generators of SU(2), respectively. The couplings $g$, $m$ and $\lambda$ are the parameters of the theory. For the results below, the bare lattice couplings take the values $\lambda=0$, $m=0$, and $g=1.342$. Since the desired correlation functions are gauge-dependent, we choose the minimal Landau gauge \cite{Maas:2011se}, also since most continuum results are in this gauge, and it is technically particularly convenient and numerically cheap. Details regarding the implementation can be found in \cite{Maas:2010nc,Maas:2012tj}.

The scale for these calculations is set such that the lightest state has a mass of 1 GeV, inspired by the scale set by the proton. For the selected lattice parameters, this corresponds to a lattice spacing of $a^{-1}=694$ MeV, a rather coarse lattice, but with considerable reach into the infrared.

\section{Propagators}

There are three two-point functions, or propagators, in the theory, all color-diagonal: The gluon propagator $D$, the ghost propagator $D_G$, and the flavor-singlet scalar propagator $D_S$. The flavor-triplet scalar propagator vanishes, as the flavor symmetry is unbroken throughout the phase diagram. While gluon and ghost propagators are only multiplicatively renormalized, the scalar propagator requires also a mass renormalization. Our renormalization scheme is to require the propagator and its derivative to coincide with a massive propagator of mass $m_r$ at $\mu=m_r$, see \cite{Maas:2010nc}.

\begin{figure}
\centering
\includegraphics[width=1.0\linewidth]{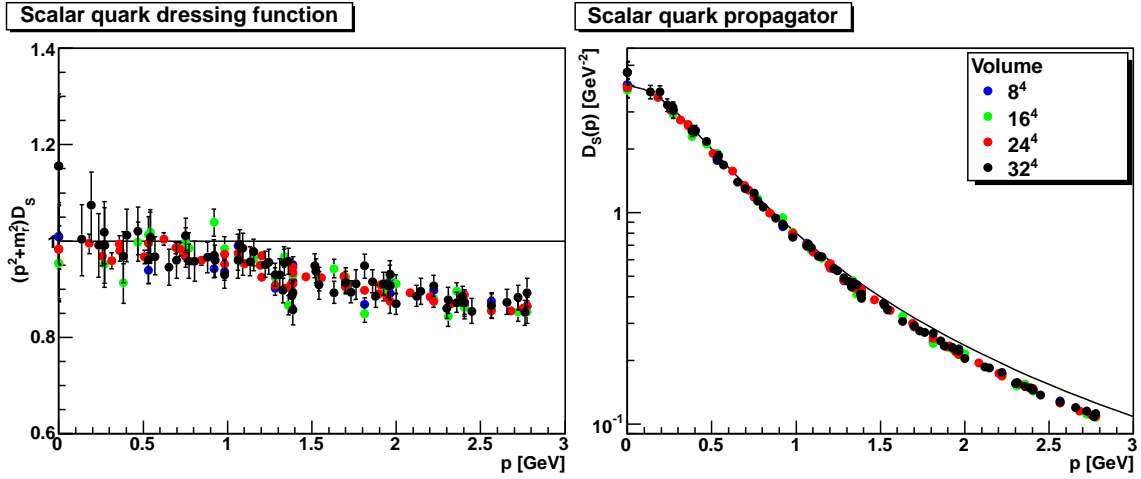}
\caption{\label{ds}Dressing function (left) and propagator(right) for the scalar quark. The line is the tree-level propagator $1/(p^2+m_r^2)$.}
\end{figure}

Once the propagator is renormalized, it can be compared with a tree-level or perturbative propagator. This is easier for the dressing functions $p^2D$, $p^2D_G$, and $(p^2+m_r)^2D_S$. Both the propagator and the dressing function of the scalar are shown for different volumes in figure \ref{ds}, where we choose $m_r=0.5$ GeV, inspired by a constituent picture. At small momenta the propagator is in good agreement to a massive propagator. At larger momenta, t decays quicker, which is most likely a perturbative loop correction. Hence it seems that for the scalar quark non-perturbative effects are small. The propagator is also very close to the corresponding quenched one \cite{Maas:2011yx}, as already observed in \cite{Maas:2010nc}. Furthermore, volume effects are small, but start to increase the smaller the renormalized mass is chosen \cite{Maas:unpublished3}.

\begin{figure}
\centering
\includegraphics[width=1.0\linewidth]{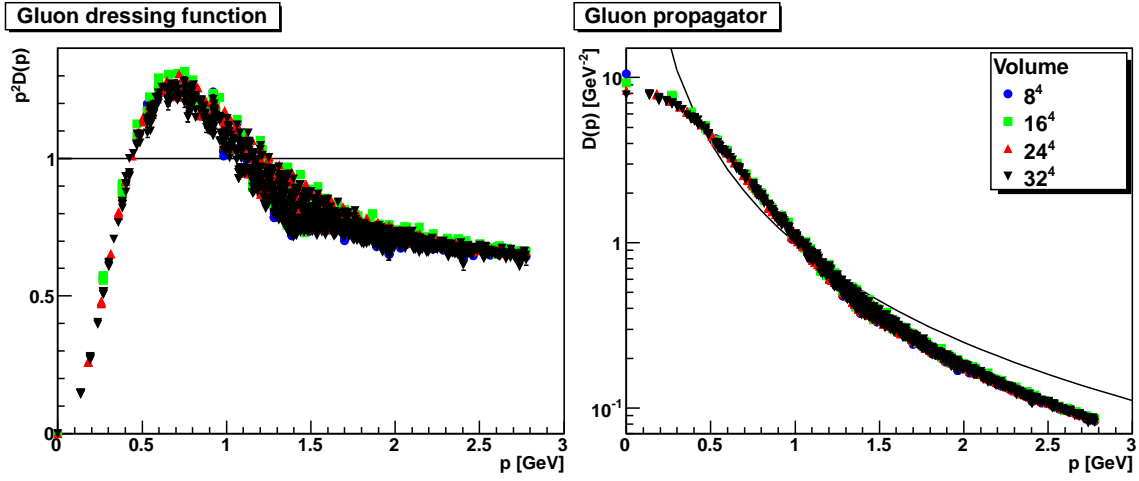}
\caption{\label{d}Dressing function (left) and propagator (right) for the gluon. The dressing function is renormalized to one at $\mu=1$ GeV. The line is the tree-level propagator $1/p^2$.}
\end{figure}

The renormalized gluon propagator is shown in figure \ref{d}. Both in the infrared and ultraviolet the deviation from tree-level is strong. The ultraviolet is likely a perturbative effect, while the infrared modification, as well as the sensitivity to finite-volume effects, is very similar to the one in the Yang-Mills case \cite{Maas:2011se,Maas:2010nc} and in the QCD case \cite{Kamleh:2007ud}. Especially, the propagator appears to exhibit a screening mass.

\begin{figure}
\centering
\includegraphics[width=1.0\linewidth]{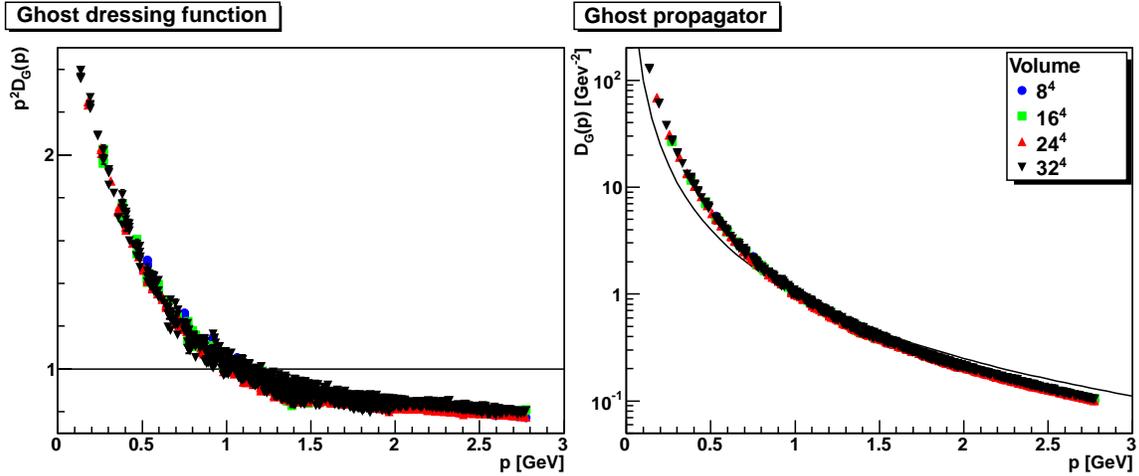}
\caption{\label{dg}Dressing function (left) and propagator (right) for the ghost. The dressing function is renormalized to one at $\mu=1$ GeV. The line is the tree-level propagator $1/p^2$.}
\end{figure}

The ghost propagator and dressing function is plotted in figure \ref{dg}. There is again a strong disagreement between the propagator and the tree-level behavior, and the ghost propagator seems to diverge. Again, the high-momentum behavior is likely dominated by perturbative corrections, while the infrared behavior is close to the one of Yang-Mills theory \cite{Maas:2011se,Maas:2010nc}. This also suggest that its dressing function is infrared finite on sufficiently large volumes, like in Yang-Mills theory \cite{Maas:2011se,Cucchieri:2008fc}, though the limited set of volumes here permits no clear statement.

Given these statistics and parameters, it is concluded that the results are very close to the quenched case \cite{Maas:2011yx}, with strong non-perturbative contributions only in the gauge sector. Furthermore, we note in passing that the scalar propagator exhibits much stronger statistical fluctuations than the gauge propagators, a nuisance becoming more severe for the vertices.

\section{Vertices}

In the Lagrangian there are 3-point and 4-point vertices. Statistical limitations make it so far not feasible to determine the 4-point vertices on the lattice, and due to the noted stronger statistical fluctuations of the scalar field compared to the gluon field, the scalar-gluon vertex is just so possible. 3-point vertices have three independent kinematic variable. Here, following \cite{Cucchieri:2006tf}, we chose two particular settings, the symmetric setting with all momenta having equal magnitude, or a setting where the momentum of (one of) the gluon leg(s) vanishes. Further kinematic choices will be made available elsewhere \cite{Maas:unpublished3}.

While the ghost-gluon vertex and the scalar-gluon vertex have only one independent transverse tensor structure, the 3-gluon vertex has four. As in \cite{Cucchieri:2006tf}, this 3-gluon vertex is projected on the tree-level tensor. We furthermore amputate and normalize the extracted functions such that the remaining function $G$ is one, up to a renormalization constant, if the full vertex would coincide with the bare one.

\begin{figure}
\centering
\includegraphics[width=1.0\linewidth]{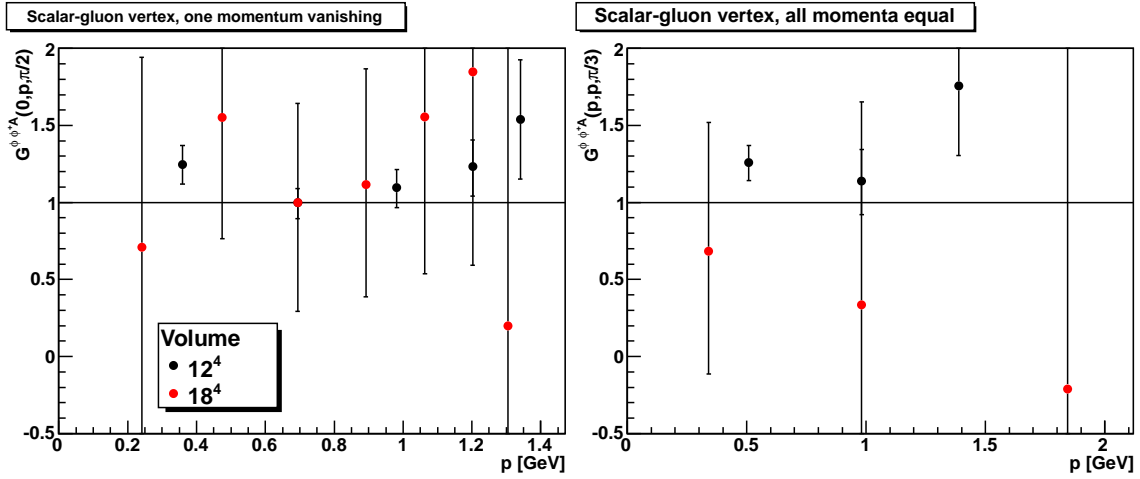}
\caption{\label{sgv}Scalar-gluon vertex. The straight line is the tree-level value.}
\end{figure}

\begin{figure}
\centering
\includegraphics[width=1.0\linewidth]{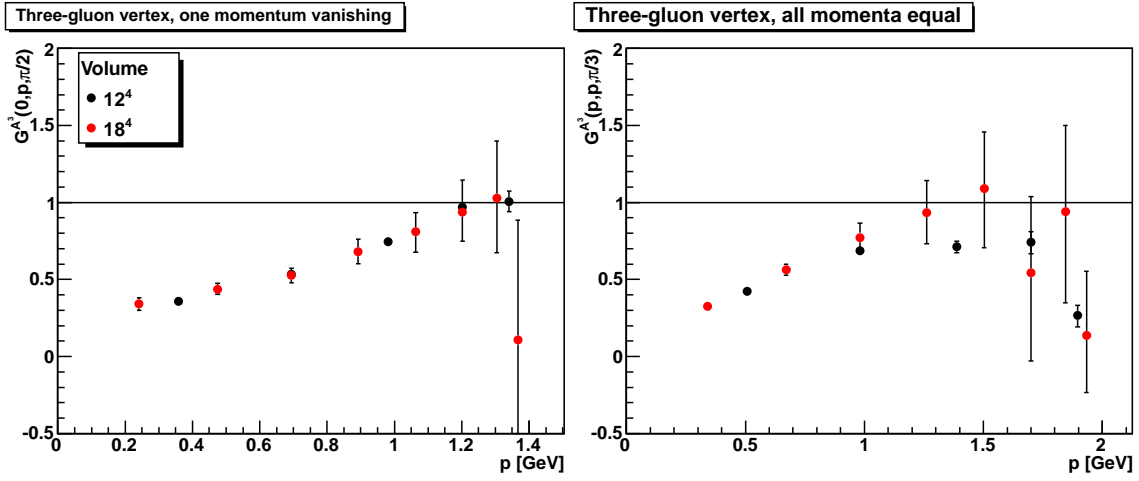}
\caption{\label{g3v}Three-gluon vertex. The straight line is the tree-level value.}
\end{figure}

\begin{figure}
\centering
\includegraphics[width=1.0\linewidth]{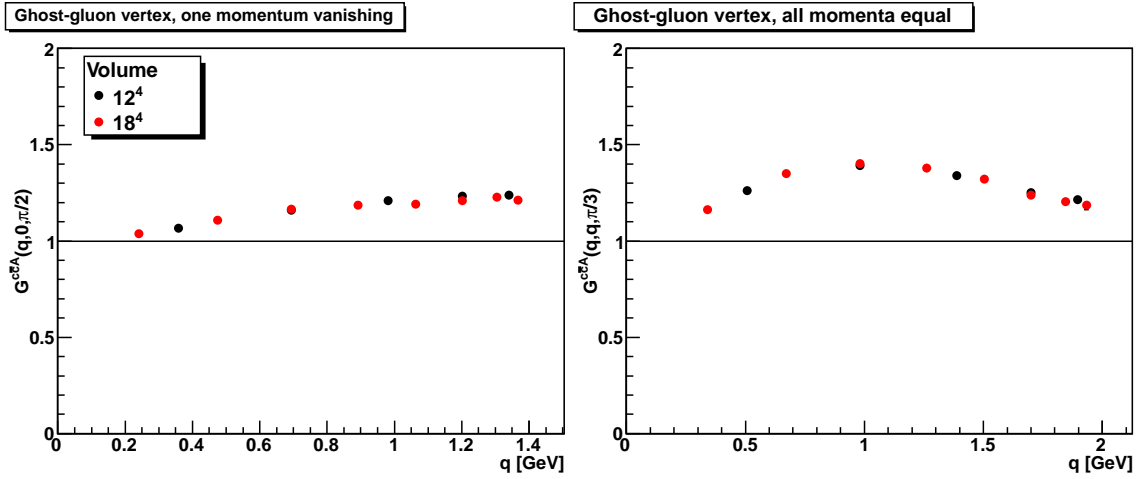}
\caption{\label{ggv}Ghost-gluon vertex. The straight line is the tree-level value.}
\end{figure}

The scalar-gluon vertex is shown in figure \ref{sgv}. The statistical errors are still large, especially for the larger volume, but the result appears to be close to tree-level, in agreement with the quenched case \cite{Maas:2011yx}. Note that a flavor-violating vertex is found to be zero \cite{Maas:unpublished3}, as expected. The three-gluon vertex and ghost-gluon vertex, shown in figures \ref{g3v} and \ref{ggv}, respectively, are better behaved, and found to be qualitatively quite similar to the Yang-Mills case \cite{Cucchieri:2006tf,Maas:2011se,Cucchieri:2008qm}, though larger volumes will be necessary to decide whether there is a zero-crossing for the 3-gluon vertex.

\section{Conclusion}

Our results so far indicate that the two-point and three-point correlation functions of scalar QCD are very close to the corresponding quenched results, similar to the case of QCD. Especially, the matter propagator and gluon-matter vertex are quite close to (renormalized) tree-level expressions, while the gauge sector is very close to the Yang-Mills case, even for the here presented case of a, at tree-level, massless scalar. In view of the discussion of \cite{Fister:2010yw,Fischer:2009tn,Alkofer:2008tt}, this suggests that the matter sector rather decouples. It furthermore implies that neither string tension nor other effects \cite{Alkofer:2008et}, which appear to be easily realized with infrared divergences in the three-point functions, seem to emerge so simply in minimal Landau gauge. This may be different in other gauges, but it could well be that in minimal Landau gauge all orders of the infinite series of correlation functions, which could contribute to such quantities, are necessary. It is not yet clear, whether this also applies to the Regge trajectory, but it constitutes a major challenge. This is especially relevant for the question to which extent minimal Landau gauge is a suitable choice for hadron spectroscopy beyond the ground states.

\bibliographystyle{bibstyle}
\bibliography{bib}

\end{document}